\pdfoutput=1
\documentclass[11pt]{article}
\usepackage[T1]{fontenc}
\usepackage{lmodern,microtype}
\usepackage[a4paper,margin=26mm]{geometry}
\usepackage{amsmath,amssymb,amsthm,mathtools}
\usepackage{xcolor}
\definecolor{ink}{HTML}{20476B}
\usepackage[colorlinks=true,linkcolor=ink,citecolor=ink,urlcolor=ink]{hyperref}
\usepackage{fancyhdr}
\allowdisplaybreaks[1]
\newtheorem{theorem}{Theorem}
\newtheorem{lemma}[theorem]{Lemma}
\newtheorem{proposition}[theorem]{Proposition}
\newtheorem{corollary}[theorem]{Corollary}
\theoremstyle{remark}
\newtheorem*{remark}{Remark}
\DeclareMathOperator{\Tr}{Tr}

\newcommand{\id}{\mathrm{id}}
\newcommand{\E}{\mathcal{E}}
\newcommand{\Alg}{\mathcal{A}}
\newcommand{\HS}{\mathrm{HS}}
\hypersetup{pdftitle={Monotonicity of skew information and its applications in quantum resource theory: Author revision},pdfauthor={Weijing Li}}
\title{\textbf{Monotonicity of skew information and its applications in quantum resource theory}\\[.5em]
{\large Expanded author revision}}
\author{Weijing Li}
\date{30 November 2026}
\begin{document}
\maketitle
\thispagestyle{fancy}
\noindent\textit{Provenance.} This author revision expands the
\href{https://arxiv.org/abs/1811.04254v4}{arXiv:1811.04254v4} preprint.
The original article was published in \emph{Quantum Information Processing}
\textbf{18}, 166 (2019), \href{https://doi.org/10.1007/s11128-019-2284-8}{DOI: 10.1007/s11128-019-2284-8}.
The additional results below belong to this author revision, not to the 2019 version of record.

\begin{abstract}
We revisit the skew information of a state relative to a quantum channel. Under the original fixed-Kraus hypotheses, we prove strong and ordinary monotonicity for \(0<p\leq2\), covering the logarithmic order and the quadratic endpoint and using an extended-valued convention for singular states. We distinguish the original generalized quantity from an affine-centered nonnegative version and describe their behavior for bistochastic reference channels. The minimum coarse-grained value is given by the conditional expectation onto the reference Kraus algebra. For pure bipartite inputs and a local bistochastic reference, the fine-grained optimum is a convex roof for \(0<p<1\); for \(1\leq p\leq2\), it is either zero or infinite according to an explicit support condition. A depolarizing example gives an exact coarse--fine gap at every order. Counterexamples identify two independent limitations: channel commutation alone is insufficient, and extending the same formula to \(p>2\) fails even under the original fixed-Kraus hypotheses.
\end{abstract}

\section{Introduction}
Wigner and Yanase introduced skew information to quantify the portion of the uncertainty of an observable attributable to its noncommutativity with a state \cite{wigner1963information,wigner1964positive,luo2003wigner}. Subsequent work examined its behavior under measurements \cite{luo2007skew} and extended it to the interaction between a state and a quantum channel \cite{luo2018coherence}. A resource theory specifies free states and transformations \cite{chitambar2018quantum,coecke2016mathematical}. Examples include entanglement \cite{horodecki2009quantum}, coherence \cite{winter2016operational,chitambar2016comparison}, reference frames \cite{gour2008resource}, and thermodynamics \cite{brandao2015reversible}. Any proposed resource measure must be assessed relative to the specific class of operations permitted by the framework.

A resource-destroying channel fixes the free states and sends every state to a free state \cite{liu2017resource,gour2017quantum}. It thereby supplies a reference channel without requiring optimization over free states \cite{kollas2018optimization}. We study the monotonicity of a generalized skew quantity with this reference held fixed. Section~\ref{sec:mono} resolves the range of orders left open in the preprint without widening the hypotheses. Section~\ref{sec:resource} treats the original applications, computes the coarse-grained minimum, and solves a pure-input fine-grained problem for local bistochastic references. Section~\ref{sec:discussion} distinguishes commutation of channels as superoperators from the stronger fixed-Kraus hypothesis. Appendices~\ref{app:commuting}--\ref{app:equality} separate the limitations concerning the operation class, order range, and equality at the endpoint.

\section{Monotonicity of skew information}\label{sec:mono}

All Hilbert spaces in the main results are finite-dimensional. Write
\(\lambda(X)=\sum_i K_i X K_i^\dagger\) for the reference channel and
\(\E(X)=\sum_n M_n X M_n^\dagger\) for the processing channel. Both are completely positive and trace preserving (CPTP). Adjoints are taken with respect to the Hilbert--Schmidt pairing. For a fixed instrument with one Kraus operator per branch, write \(q_n=\Tr M_n\rho M_n^\dagger\) and \(\rho_n=M_n\rho M_n^\dagger/q_n\) when \(q_n>0\). We omit zero-probability branches.

\subsection{The original skew information and the fixed-Kraus hypotheses}
For a Hermitian \(H\), the Wigner--Yanase convention is
\(I_{\rm WY}(\rho,H)=-\tfrac12\Tr[\sqrt{\rho},H]^2\).
Following the preprint for general \(K\), set
\begin{equation}\label{eq:commutator}
 I(\rho,K)=\|[\sqrt{\rho},K]\|_{\HS}^2,\qquad
 I(\rho,\lambda)=\sum_i I(\rho,K_i).
\end{equation}
For a single Hermitian \(K\), this convention is twice \(I_{\rm WY}\).

\begin{lemma}[One-sided Schwarz equality]\label{lem:one}
For a channel \(\E\) as above, \(\E^\dagger(K)=K\) and
\(\E^\dagger(K^\dagger K)=K^\dagger K\) are equivalent to
\([K,M_n]=0\) for every \(n\). Likewise,
\(\E^\dagger(K)=K\) and \(\E^\dagger(KK^\dagger)=KK^\dagger\) are equivalent to
\([K^\dagger,M_n]=0\) for every \(n\).
\end{lemma}
\begin{proof}
If \(K\) is fixed, expanding \(\sum_n(KM_n-M_nK)^\dagger(KM_n-M_nK)\) gives
\(\E^\dagger(K^\dagger K)-K^\dagger K\).
Its vanishing is equivalent to the vanishing of each summand.
Conversely, \([K,M_n]=0\) and \(\sum_n M_n^\dagger M_n=I\) give the two fixed-point identities. Apply the same argument to \(K^\dagger\) for the other assertion.
\end{proof}

\begin{lemma}[Reduction of the hypotheses]\label{lem:both}
Suppose that for each \(i\)
\begin{equation}\label{eq:hyp}
 \E^\dagger(K_i)=K_i,\qquad
 \E^\dagger(K_iK_i^\dagger)=K_iK_i^\dagger .
\end{equation}
Then \(\E^\dagger(K_i^\dagger K_i)=K_i^\dagger K_i\) for every \(i\). In particular all \(M_n\) commute with all \(K_i\) and \(K_i^\dagger\).
\end{lemma}
\begin{proof}
The Schwarz inequality for the unital CP map \(\E^\dagger\) gives
\(D_i=\E^\dagger(K_i^\dagger K_i)-K_i^\dagger K_i\geq0\).
Since \(\sum_i K_i^\dagger K_i=I\) and \(\E^\dagger(I)=I\), we have \(\sum_iD_i=0\), so each \(D_i=0\). Both commutation relations now follow from Lemma~\ref{lem:one}.
\end{proof}

\setcounter{theorem}{3}
\begin{theorem}[Commutator skew information; original Theorem~4]\label{thm:four}
Under the original assumptions that \(\E^\dagger\) fixes \(K_i\), \(K_i^\dagger K_i\), and \(K_iK_i^\dagger\) for all \(i\),
\[
 I(\rho,\lambda)\ \geq\ \sum_{n:q_n>0}q_n I(\rho_n,\lambda)\
 \geq\ I(\E(\rho),\lambda).
\]
For the reference Kraus family of a channel, the first and third fixed-point identities already suffice, by Lemma~\ref{lem:both}.
\end{theorem}
The theorem follows from the order-\(\tfrac12\) case of Theorem~\ref{thm:five} and the affine relation in \eqref{eq:center}. The Hermitian-observable statement discussed in the preprint also follows directly: if \(\E^\dagger(H)=H\) and \(\E^\dagger(H^2)=H^2\), then the \(M_n\) commute with \(H\). Apply the recorded-outcome argument below to the single insertion \(H\) and the kernel \((\sqrt{x}-1)^2\), whose associated quasi-entropy evaluated at \((\rho,\rho)\) is \(\|[\sqrt{\rho},H]\|_{\HS}^2\). This gives strong monotonicity of \(I_{\rm WY}(\rho,H)\), including the case treated by Luo and Zhang \cite{luo2007skew}. The flag construction requires no additional dimensional hypothesis.

\subsection{The generalized skew quantity}
Following the unified relative-entropy convention \cite{jenvcova2010unified}, define
\begin{equation}\label{eq:gp}
 g_p(x)=\begin{cases}
 (x^p-x)/[p(p-1)],&p\ne1,\\
 x\log x,&p=1,
 \end{cases}\qquad 0<p\leq2.
\end{equation}
For positive definite \(A,B\), write \(L_A(X)=AX\) and \(R_B(X)=XB\), and define
\begin{equation}\label{eq:quasi}
 S_{g_p}^{K}(A\|B)
  =\big\langle K\sqrt B,\,g_p(L_A R_B^{-1})(K\sqrt B)\big\rangle_{\HS}.
\end{equation}
For instance \(S_{g_1}^I(A\|B)=\Tr A(\log A-\log B)\).
For positive definite \(A,B\), the order-two expression is
\(\tfrac12[\Tr(K^\dagger A^2KB^{-1})-\Tr(K^\dagger AK)]\);
the order of the factors matters.

For positive semidefinite \(A=\sum_a aP_a\), \(B=\sum_b bQ_b\), the extended spectral definition is
\begin{equation}\label{eq:spectral}
 S_{g_p}^{K}(A\|B)=\sum_{a,b} \|P_a KQ_b\|_{\HS}^2\,\mathcal P_{g_p}(a,b),
 \qquad
 \mathcal P_f(a,b)=b f(a/b)\quad(b>0).
\end{equation}
Here \(f(0)=0\) for \(f=g_p\), \(\mathcal P_f(a,0)=a f'(\infty)\) for \(a>0\), and \(\mathcal P_f(0,0)=0\), with \(f'(\infty)=\lim_{x\to\infty}f(x)/x\). Zero coefficients do not contribute. Equivalently, the prescription is obtained by evaluating the expression at \(A+\varepsilon I,B+\varepsilon I\) and taking \(\varepsilon\downarrow0\); it never yields \(-\infty\). We retain the preprint's definition
\begin{equation}\label{eq:ip}
 I_p(\rho,\lambda)=\sum_i S_{g_p}^{K_i}(\rho\|\rho).
\end{equation}
The quantity \(I_p\) is independent of the Kraus representation: for any fixed \(a,b\),
\(\sum_i\|P_a K_i Q_b\|_{\HS}^2=\Tr[P_a\lambda(Q_b)]\).
For \(1\leq p\leq2\) the finiteness criterion is
\(P K_i(I-P)=0\) for all \(i\), where \(P\) is the support projection of \(\rho\); for \(0<p<1\) all terms are finite.

\subsection{Extension of Theorem 5 to the endpoint and logarithmic order}
\begin{theorem}[Revised Theorem 5]\label{thm:five}
Under \eqref{eq:hyp}, for every density operator \(\rho\) and \(0<p\leq2\),
\begin{equation}\label{eq:five}
 I_p(\rho,\lambda)\ \geq\
 \sum_{n:q_n>0}q_n I_p(\rho_n,\lambda)\
 \geq\ I_p(\E(\rho),\lambda).
\end{equation}
The function \(I_p(\,\cdot\,,\lambda)\) is convex, with the convention \eqref{eq:spectral}. The first inequality also holds for coarser outcomes obtained by grouping the Kraus labels.
\end{theorem}

\begin{proof}
\textit{Step 1: joint convexity, including the endpoints.}
For \(t>0\) let \(h_t(x)=(x-1)^2/(x+t)\).
In the joint eigenbasis of \(L_A,R_B\), its quasi-entropy is
\begin{equation}\label{eq:frac}
 S_{h_t}^K(A\|B)=
 \langle AK-KB,\,(L_A+tR_B)^{-1}(AK-KB)\rangle_{\HS}.
\end{equation}
For \(A,B>0\) this follows by inserting the spectral weights:
the coefficient before \(\|P_aKQ_b\|_{\HS}^2\) is
\((a-b)^2/(a+tb)\). The matrix-fractional form equals
\[
 \sup_Y\{2\operatorname{Re}\langle Y,AK-KB\rangle_{\HS}
               -\langle Y,(L_A+tR_B)Y\rangle_{\HS}\}.
\]
For each fixed \(Y\) the expression inside the supremum is affine in \((A,B)\), so \eqref{eq:frac} is jointly convex. For \(0<p<2\), including \(p=1\), \(g_p''(x)=x^{p-2}\) and \(g_p'(1)=1/p\). The identity
\begin{equation}\label{eq:beta}
 g_p(x)=\frac{x-1}{p}
 +\frac{1}{2\,\mathrm B(p+1,2-p)}
       \int_0^\infty\frac{t^p}{(1+t)^2}\,h_t(x)\,dt
\end{equation}
follows by differentiating twice, using \(h_t''(x)=2(1+t)^2/(x+t)^3\),
and matching value and derivative at \(x=1\). Here \(\mathrm B\) is the Euler beta function. The integrals converge at both ends for positive \(x\) and \(0<p<2\); differentiation is justified locally uniformly for \(x>0\).
At \(p=2\), \(g_2(x)=\tfrac12(x-1)+\tfrac12(x-1)^2\).
The squared kernel has the same matrix-fractional form \eqref{eq:frac}, with \(L_A+tR_B\) replaced by \(R_B\). The affine term in the representation also contributes an affine function of \((A,B)\); this proves joint convexity on positive pairs for the entire interval. For semidefinite pairs, add the same \(\varepsilon I\) to both matrices and take the spectral limit \eqref{eq:spectral}. This establishes extended joint convexity, including for singular \(\rho\) and singular branches.

\smallskip
\textit{Step 2: intertwining through an outcome-recording isometry.}
Use the isometry
\[
 V\psi=\sum_n M_n\psi\otimes |n\rangle,\qquad
 \sigma=V\rho V^\dagger,\qquad \widehat K_i=K_i\otimes I_{\rm flag}.
\]
Lemma~\ref{lem:both} gives both \(\widehat K_i V=VK_i\) and
\(\widehat K_i^\dagger V=VK_i^\dagger\). Thus the range of \(V\) is reducing for \(\widehat K_i\). On this range, \(\widehat K_i\) and \(\sigma\) are unitarily identified with \(K_i\) and \(\rho\), respectively. On its orthogonal complement \(\sigma\) vanishes and there are no cross terms in \eqref{eq:spectral}. Consequently
\begin{equation}\label{eq:intertw}
 S_{g_p}^{\widehat K_i}(\sigma\|\sigma)
 =S_{g_p}^{K_i}(\rho\|\rho).
\end{equation}

\smallskip
\textit{Step 3: dephase the flag.}
The pinching \(\mathcal P\) in the flag basis is the average of conjugations by diagonal phase unitaries on the flag. They commute with \(\widehat K_i\), so each conjugation leaves its quasi-entropy unchanged. By joint convexity,
\begin{equation}\label{eq:pinch}
 S_{g_p}^{\widehat K_i}(\mathcal P\sigma\|\mathcal P\sigma)
  \leq S_{g_p}^{\widehat K_i}(\sigma\|\sigma),\qquad
 \mathcal P\sigma=\sum_n M_n\rho M_n^\dagger\otimes |n\rangle\langle n|.
\end{equation}
Since the insertion \(\widehat K_i\) is block diagonal, the left side sums the block quasi-entropies. Homogeneity in the pair \((A,B)\) gives the block weight \(q_n\) for each occupied block. Summing \eqref{eq:intertw}--\eqref{eq:pinch} proves the first inequality of \eqref{eq:five}. Convexity applied to \(\E(\rho)=\sum_n q_n\rho_n\) proves the second and also proves the coarse-outcome assertion by convexity within each group.
\end{proof}

\begin{remark}[Relation to general data processing]
The ordinary inequality is also a direct instance of the operator-convex quasi-entropy data-processing theorem on the multiplicative domain \cite[Theorem~2.16]{hiai2023equality}. Applying that theorem to the recorded-outcome channel and insertions \(K_i\otimes I_{\rm flag}\) gives the flagged inequality. The proof above gives the finite-dimensional calculation explicitly and does not assert priority over the general theorem. See also \cite{hiai2012quasi,vershynina2018quantum} for background on quasi-entropies and partial trace.
\end{remark}

\subsection{Nonnegativity and normalization}\label{sec:normalization}
Trace preservation of the reference alone is not enough to make the original \(I_p\) nonnegative (Appendix~\ref{app:checks}). The centered kernel \(g_p(x)-(x-1)/p\) is nonnegative and vanishes only at \(x=1\); its spectral perspective gives the centered quantity
\begin{equation}\label{eq:center}
 J_p(\rho,\lambda):=I_p(\rho,\lambda)
     +\frac{\Tr[\rho(I-\lambda(I))]}{p},
 \qquad J_{1/2}(\rho,\lambda)=2I(\rho,\lambda).
\end{equation}
The identity at \(p=\tfrac12\) follows because the centered kernel equals \(2(\sqrt{x}-1)^2\). In \eqref{eq:center}, the affine term is the difference of the two spectral first moments:
\(\sum_i\Tr(\rho K_iK_i^\dagger)=\Tr\rho\lambda(I)\) and
\(\sum_i\Tr(\rho K_i^\dagger K_i)=1\).
It is preserved under \(\E\) and under averaging over all branches, so \eqref{eq:five} applies unchanged to \(J_p\). This proves Theorem~\ref{thm:four}.

For every reference channel, \(J_p\geq0\). Moreover, \(J_p=0\) if and only if \([\rho,K_i]=0\) for every \(i\), because the centered perspective is strictly positive between distinct spectral values, including terms coupling the support to the kernel, where it may be infinite. For a bistochastic reference channel, \(J_p=I_p\). In that case \(\lambda(\rho)=\rho\) holds exactly when the commutators vanish: one implication follows from
\(\sum_i\|[\rho,K_i]\|_{\HS}^2=2\Tr\rho^2-2\Tr[\rho\lambda(\rho)]\);
the other uses \(\lambda(I)=I\). For a nonunital reference channel we do not identify the zero set of \(J_p\) with the fixed states of \(\lambda\).

\section{Applications in the resource framework}\label{sec:resource}
A resource-destroying map fixes every free state and sends every input to a free state \cite{liu2017resource,gour2017quantum}. The following two examples are bistochastic, so the quantities above are monotone, nonnegative, and faithful.

\paragraph{Asymmetry.}
For a finite group \(G\) with unitary representation \(U_g\), define the reference channel
\[\lambda(\rho)=|G|^{-1}\sum_g U_g\rho U_g^\dagger.\]
The free states are those commuting with every \(U_g\) \cite{marvian2012symmetry,marvian2013theory}. The fixed-observable hypothesis \(\E^\dagger(U_g)=U_g\) implies \eqref{eq:hyp} because \(U_gU_g^\dagger=I\); thus Theorem~\ref{thm:five} applies for all \(0<p\leq2\). For a compact group with a continuous finite-dimensional representation, the channel average is the Haar twirl. If the adjoint fixes every \(U_g\), the processing Kraus operators commute with all of them. The same conclusion therefore holds; in finite dimension, the twirl can be realized as a finite convex combination of group conjugations.

\paragraph{Coherence.}
For the dephasing reference \(\Delta(\rho)=\sum_iP_i\rho P_i\), \(P_i=|i\rangle\langle i|\), the fixed-Kraus assumption is \(\E^\dagger(P_i)=P_i\) for every \(i\). Consequently, the processing Kraus operators are diagonal in this basis. Hence \(I_p(\rho,\Delta)\) is strongly monotone for this specified class; this does not by itself imply strong monotonicity under all incoherent operations \cite{winter2016operational,streltsov2017colloquium,marvian2016quantify}. At \(p=\tfrac12\), the definition reduces to
\(4[1-\sum_i\langle i|\sqrt\rho|i\rangle^2]\), proportional to Yu's coherence measure \cite{yu2017quantum}.

\subsection{The minimum under the stated operations}
Let \(\Alg\) be the unital \(*\)-algebra generated by all the \(K_i\), and \(\Alg'\) its commutant. By Lemma~\ref{lem:both}, \eqref{eq:hyp} is equivalent to \(M_n\in\Alg'\) for every \(n\). The converse follows by commuting \(K_i\) and \(K_i^\dagger\) through the adjoint Kraus sum. In particular \(\E^\dagger\) fixes \(\Alg\) pointwise. Both the criterion and the algebra are independent of the choice of Kraus representations, since the linear span of the Kraus operators is determined by the support of the Choi matrix.

\begin{corollary}[Attainable minimum]\label{cor:floor}
Let \(E_{\Alg}\) denote the trace-preserving conditional expectation onto \(\Alg\). Then the minimum of \(I_p(\E(\rho),\lambda)\) over channels satisfying \eqref{eq:hyp} equals \(I_p(E_{\Alg}\rho,\lambda)\). It is also the minimum average when instrument outcomes may combine multiple Kraus operators. The same assertions hold for \(J_p\).
\end{corollary}
\begin{proof}
Averaging conjugations by the unitaries in \(\Alg'\) gives \(E_{\Alg}\), so \(E_{\Alg}\) is admissible (this finite-dimensional twirl has a finite convex realization). Since every admissible adjoint fixes \(\Alg\) pointwise, testing against every \(A\in\Alg\) gives \(E_{\Alg}\E(\rho)=E_{\Alg}\rho\). Applying the ordinary inequality to the composition \(E_{\Alg}\circ\E\) proves the lower bound, and choosing \(\E=E_{\Alg}\) attains it. The average is bounded below by the nonselective value by convexity, and the bound is attained by the single coarse outcome \(E_{\Alg}\).
\end{proof}

In the usual block decomposition \cite{holbrook2004noiseless}
\[
 \Alg=\bigoplus_\alpha\bigl(M_{d_\alpha}\otimes I_{m_\alpha}\bigr),\qquad
 E_{\Alg}(\rho)=\bigoplus_\alpha
     \Tr_{\mathbb C^{m_\alpha}}(P_\alpha\rho P_\alpha)
                   \otimes \frac{I_{m_\alpha}}{m_\alpha},
\]
where \(P_\alpha\) projects onto that summand. The centered minimum is zero precisely when each reduced block is scalar on its \(d_\alpha\)-dimensional factor, equivalently \(E_{\Alg}\rho\) lies in the center of \(\Alg\). Thus a single admissible channel can make \(J_p=0\) on every input if and only if \(\Alg\) is abelian. If each individual Kraus outcome must be retained, attainability of the average floor can fail (Appendix~\ref{app:checks}).

\subsection{What changes when every Kraus outcome is kept}\label{sec:fine}
The minimum in Corollary~\ref{cor:floor} allows a channel to be recorded as one outcome, even when that outcome has several Kraus operators. This is not the same as minimizing the average over instruments that keep every single-Kraus outcome. For pure bipartite inputs and a local bistochastic reference, the latter problem reduces to a convex roof.

Let \(\Gamma(X)=\sum_i L_i X L_i^\dagger\) be bistochastic on \(A\), and take \(\lambda=\Gamma\otimes\id_B\). For an input \(|\psi\rangle_{AB}\) put \(\eta=\Tr_B|\psi\rangle\langle\psi|\). We restrict processing to \(B\), so it satisfies \eqref{eq:hyp}. For a pure state \(|u\rangle_A\) define
\[
 c_\Gamma(u)=1-\sum_i|\langle u|L_i|u\rangle|^2,\qquad
 C_\Gamma(\eta)=\min_{\sum_jq_j|u_j\rangle\langle u_j|=\eta}
                      \sum_j q_j c_\Gamma(u_j).
\]
Let \(F_p(\psi;\Gamma)\) be the infimum of the average \(I_p\) over local instruments on \(B\) with one Kraus operator per reported outcome. For a tuple \(z=(z_i)_i\), write \(H_z=\bigcap_i\ker(L_i-z_iI)\) and index only the nonzero subspaces. These are orthogonal reducing subspaces, as justified below.

\begin{theorem}[Fine-grained optimum for pure inputs]\label{thm:roof}
For \(0<p<1\), \(F_p(\psi;\Gamma)=C_\Gamma(\eta)/[p(1-p)]\), attained by a finite instrument. For \(1\leq p\leq2\), this optimum is zero if and only if \(\eta\) is block diagonal on \(\bigoplus_z H_z\) and has no support on its orthogonal complement; otherwise, it is \(+\infty\). When grouping Kraus operators is permitted, the optimal average over local processing is instead \(I_p(\eta,\Gamma)\).
\end{theorem}

\begin{proof}
Each occupied single-Kraus branch is pure. Write its projection as \(P\) and its \(A\)-marginal as \(\omega\). For a pure state the only nonzero spectral contribution of \(g_p\) is proportional to the recession slope: the term with numerator eigenvalue \(0\) and denominator eigenvalue \(1\) vanishes because \(g_p(0)=0\), and the term with both equal to \(1\) vanishes as well. Its coefficient is
\[
 \sum_i\Tr\!\left[P(L_i\otimes I)(I-P)(L_i^\dagger\otimes I)\right]
   =1-\sum_i|\Tr(\omega L_i)|^2.
\]
Here unitality supplies the first term \(1\). The slope is \(1/[p(1-p)]\) when \(0<p<1\) and \(+\infty\) for \(p\geq1\).

The \(A\)-marginals of the branches average to \(\eta\). The function \(1-\sum_i|\Tr(\omega L_i)|^2\) is concave. Refining each mixed marginal into a pure-state ensemble does not increase its average cost; hence the convex roof lower bounds every fine-grained average for \(p<1\). Conversely, the Hughston--Jozsa--Wootters ensemble theorem \cite{hjw1993} realizes any finite pure-state decomposition of \(\eta\) by a rank-one POVM on the purifying system \(B\); choosing one rank-one Kraus operator per outcome then yields product branches with exactly those \(A\)-states. The convex roof minimum exists with a finite ensemble by compactness of the convex hull of the graph of \(c_\Gamma\) on pure states.

For \(p\geq1\), the coefficient is zero precisely when the pure branch commutes with every \(L_i\otimes I\): its vanishing is equivalent to \(\lambda(P)=P\), since \(\Tr P\lambda(P)=1\) and \(\lambda(P)\) is a state. The bistochastic fixed-point argument in Section~\ref{sec:normalization} then applies. In particular, \((L_i\otimes I)|\psi_{\rm branch}\rangle=z_i|\psi_{\rm branch}\rangle\), or equivalently the Schmidt support on \(A\) is in one \(H_z\). To justify the claim about \(H_z\), a unit vector \(u\) with \(L_i u=z_i u\) satisfies \(\sum_i|z_i|^2=1\) by trace preservation. The coefficient for the pure state \(u\) vanishes because \(\Tr(|u\rangle\langle u|\Gamma(|u\rangle\langle u|))=\sum_i|z_i|^2=1\). Thus \(|u\rangle\langle u|\) commutes with every \(L_i\); taking adjoints shows that \(u\) is also an eigenvector of \(L_i^\dagger\) with eigenvalue \(\overline z_i\). This makes \(H_z\) reducing, and two different tuples give orthogonal subspaces by taking matrix elements of an operator whose eigenvalues differ. Consequently all occupied branches have zero cost only if \(\eta\) is a mixture within the stated blocks. If \(\eta\) has that form, a pure-state decomposition within the blocks followed by the same ensemble realization attains zero. Otherwise at least one occupied branch has infinite cost for every instrument.

Finally, depolarizing \(B\) after any local processing produces the product \(\eta\otimes I_B/\dim B\), which has value \(I_p(\eta,\Gamma)\) by the spectral formula. Applying Theorem~\ref{thm:five} to that extra depolarization lower bounds every nonselective output, and convexity lower bounds every average by its nonselective value. A single coarse outcome applying the depolarization attains the bound.
\end{proof}

\paragraph{Exact gap for a depolarizing reference.}
If \(\Gamma(X)=\Tr(X)\,I_A/d\) and \(d=\dim A\geq2\), then \(c_\Gamma(u)=1-1/d\) for every pure \(u\). The Kraus algebra on \(A\) is the full matrix algebra, so every local channel on \(B\) is admissible and there are no common scalar subspaces \(H_z\). In particular, for the maximally entangled state of two \(d\)-dimensional systems,
\begin{equation}\label{eq:exactfine}
  \min_{\rm coarse}\sum_jq_j I_p(\rho_j,\lambda)=0,\qquad
  \min_{\rm fine}\sum_jq_j I_p(\rho_j,\lambda)=
  \begin{cases}
      (1-1/d)/[p(1-p)],&0<p<1,\\
      +\infty,&1\leq p\leq2 .
  \end{cases}
\end{equation}
For \(p<1\), measuring the Schmidt basis of \(B\) already attains the fine minimum. Thus the distinction between discarding a record and retaining each Kraus outcome changes the optimum throughout the range, not just at orders with infinite values.

\paragraph{Qubit dephasing.}
As a second explicit application, let \(\Gamma\) dephase a qubit and write
\(z=\eta_{01}\). A pure qubit with coherence \(z_j\) has \(c_\Gamma=2|z_j|^2\).
Convexity of the squared modulus gives \(C_\Gamma(\eta)\geq2|z|^2\).
Conversely, the two populations \(r_\pm=(1\pm\sqrt{1-4|z|^2})/2\) both give pure states with the prescribed coherence \(z\), and
\(\eta_{00}\) lies between them by positivity of \(\eta\). A mixture of
these pure states reproduces \(\eta\), giving \(C_\Gamma(\eta)=2|z|^2\).
Thus the fine optimum is \(2|z|^2/[p(1-p)]\) for \(0<p<1\).
For \(1\leq p\leq2\) it is zero if \(z=0\) and infinite otherwise.

\section{Discussion}\label{sec:discussion}
The extension to \(1\leq p\leq2\) uses the same hypotheses as the preprint's Theorem~5. Joint convexity and the recorded-outcome construction enable the treatment of the operator-convex orders, including \(g_1(x)=x\log x\) and \(g_2(x)=(x^2-x)/2\). This result does not claim that a general nonunital resource-destroying map yields a nonnegative measure with the same zeros as its fixed states.

The choice of operation class remains substantive. In particular, replacing \eqref{eq:hyp} by \(\E\lambda=\lambda\E\) as a relation between superoperators is insufficient, even at \(p=1\) for a bistochastic reference; Appendix~\ref{app:commuting} gives a classical counterexample based on a reversible Markov semigroup. By contrast, mixing unitary conjugations whose superoperators commute with the reference channel does imply ordinary monotonicity: each conjugation preserves \(I_p\), by unitary covariance of the spectral formula and Kraus-representation independence, and convexity applies to the mixture. The order range and equality behavior are separate questions. Appendix~\ref{app:orders} shows, within the original hypotheses, that the order bound \(p\leq2\) is sharp for the algebraic continuation. Appendix~\ref{app:equality} gives a faithful case with equality at \(p=2\) and strict loss throughout the interior. Identifying broader operation classes and comparing the dephasing-based quantities with Tsallis-based coherence measures \cite{zhao2018coherence} remain natural directions for future work.

\paragraph{Acknowledgments in the original article.}
The original article acknowledges Shunlong Luo, Shaoming Fei, and Wei Xie for discussions. These acknowledgments do not imply endorsement of this revision.

\paragraph{Use of generative AI.}
Generative AI tools assisted with exploratory derivations, calculation checks, and language editing during preparation of this expanded revision. No AI tool is an author.

\appendix
\section{Concrete checks of the monotonicity and normalization}\label{app:checks}
\paragraph{Recorded versus discarded outcomes.}
Let \(\rho=(I+X/2)/2\), take \(\lambda=\Delta\) in the \(Z\) basis, and let \(\E\) apply either \(I\) or \(Z\), with probability \(1/2\) each. Both Kraus operators satisfy \eqref{eq:hyp}. The two eigenvalues of \(\rho\) are \(3/4,1/4\); in the \(X\)-eigenbasis the squared entries of each insertion \(P_0,P_1\) are \(1/4\). Thus
\[
 I_p(\rho,\Delta)=\tfrac18 g_p(3)+\tfrac38 g_p(1/3),\qquad
 I_1(\rho,\Delta)=\tfrac14\log 3,\qquad I_2(\rho,\Delta)=\tfrac13.
\]
Each outcome preserves the initial value by invariance under the corresponding diagonal unitary. If the flag is discarded, the output is \(I/2\) and \(I_p\) is zero. This checks the endpoint, the logarithmic order, and the distinction between the two inequalities of \eqref{eq:five}.

\paragraph{Why the uncentered expression can be negative.}
For \(\rho=\mathrm{diag}(1/4,3/4)\), let \(\lambda\) reset to \(|0\rangle\) with Kraus operators \(|0\rangle\langle0|,|0\rangle\langle1|\).
Equation \eqref{eq:spectral} gives \(I_p=\tfrac34g_p(1/3)<0\). It equals \(-(\log 3)/4\) at \(p=1\) and \(-1/12\) at \(p=2\). This does not contradict Theorem~\ref{thm:five}. More generally, for a faithful state with eigenvalues \(r_b\), the coefficient \(\sum_i| (K_i)_{ab}|^2\) is column stochastic. Conversely every column-stochastic matrix \(T\) is realized by \(\sqrt{T_{ab}}|a\rangle\langle b|\). Hence the smallest uncentered value over all reference channels at fixed \(\rho\) is
\(\sum_b r_b\min_a g_p(r_a/r_b)\).
For \(r=(3/5,3/10,1/10)\) at \(p=2\) this is \(-13/120\), attained by transitions \(1\to2\), \(2\to3\), \(3\to3\), with contributions \(-3/40,-1/30,0\).

\paragraph{Attainable minima and fine-graining.}
If \(\lambda\) depolarizes the first of two qubits, then \(\Alg=M_2\otimes I\) and \(E_{\Alg}\) depolarizes only the second. The states \((I+Z\otimes X/2)/4\) and \((I+Z\otimes I/2)/4\) have the same initial \(I_p\), equal to \(\tfrac18g_p(3)+\tfrac38g_p(1/3)\), by inspecting the two Pauli Kraus operators that flip the \(Z\)-eigenspaces of the first qubit. The floor of the first state is zero, whereas the second state's floor is its initial value. For a pure Bell input the coarse floor is zero. Each nonzero single-Kraus branch is pure, however, and a rank-one projection cannot commute with \(M_2\otimes I\). At least one reference Pauli Kraus operator then has an occupied-row/kernel-column term by Hermiticity. The fine-grained average is \(+\infty\) for \(1\leq p\leq2\).

\section{Why commuting channels are not sufficient}\label{app:commuting}
Take the unweighted star with one center, one high-weight leaf and \(32\) low-weight leaves. Let \(L\) be its negative graph Laplacian, \(T=I+L/33\), and \(\lambda(X)=\mathrm{diag}(T\,\mathrm{diag}\,X)\). This is a primitive bistochastic channel: \(T\) is symmetric and irreducible, and the positive leaf self-loops imply aperiodicity; the channel removes off-diagonal entries. The processing channel \(\E_t=e^{t(\lambda-\id)}\) is CPTP by its Poisson mixture representation and commutes with \(\lambda\). For diagonal input let \(x=(1,41,1/4,\ldots,1/4)\), \(Z_0=\sum_a x_a=50\) and \(\rho=\mathrm{diag}(x)/Z_0\). The diagonal masses evolve as \(x(t)=e^{tL/33}x\), and by \eqref{eq:spectral} the order-one value is the sum over unordered edges:
\[
 I_1(\rho_t,\lambda)
 =\frac{1}{50\cdot33}\sum_{\{a,b\}}(x_a(t)-x_b(t))
                       \log(x_a(t)/x_b(t)).
\]
Here \(Lx\) takes the values \(16,-40,3/4\) on the center, high leaf, and each low leaf. Differentiating the displayed edge sum at zero gives
\[
 I_1'(\rho_t,\lambda)\big|_{t=0}
 =\frac{-56\log41+488\log4-328-1600/41}{50\cdot33^2}>0.
\]
The sign already follows from \(\log41<4\), \(\log4>4/3\) and \(1600/41<40\). Thus \(\E_t\) increases the generalized skew information for sufficiently small \(t>0\). This example is outside the assumptions of Theorem~\ref{thm:five}.

\section{Sharpness of the order range}\label{app:orders}
The following statement concerns the same algebraic \(g_p\) formula formally continued to \(p>2\), not a redefinition of the quantity on the interval of Theorem~\ref{thm:five}.

\begin{proposition}[Failure at every fixed \(p>2\)]\label{prop:sharp}
For each fixed real \(p>2\), the continued formula fails both ordinary and strong monotonicity under the hypotheses \eqref{eq:hyp}, even for a unitary reference and a faithful input.
\end{proposition}
\begin{proof}
Take strictly positive \(2\times2\) matrices \(A\pm\varepsilon H\) and \(B\). Put \(T=\Tr(A+B)\), and let the two classical flag blocks of the input be
\[
  \rho=\frac{[(A+\varepsilon H)\oplus B]\oplus[(A-\varepsilon H)\oplus B]}{2T}.
\]
Within each flag block the reference unitary \(K\) exchanges the two two-dimensional sectors. Erasing the flag and replacing it by \(I/2\) is admissible. Define
\(F_p(X,B)=\Tr(X^p B^{1-p})+\Tr(B^p X^{1-p})\).
The input minus output \(I_p\) is the midpoint difference of \(F_p\) in its first argument divided by \(T p(p-1)>0\).

Fix \(H=\left(\begin{smallmatrix}1&h\\h&0\end{smallmatrix}\right)\), \(0<h^2<(p-2)/2\), and \(A=\mathrm{diag}(1,s)\). The coefficient of \(\varepsilon^2\) in \(( (A+\varepsilon H)^p+(A-\varepsilon H)^p )/2-A^p\) tends as \(s\downarrow0\) to
\[
 \begin{pmatrix}
 p(p-1)/2+(p-1)h^2 &(p-1)h\\
 (p-1)h&h^2
 \end{pmatrix}.
\]
Indeed, the matrix second derivative coefficient has entry \(\sum_k f[a_i,a_k,a_j]H_{ik}H_{kj}\), with \(f(x)=x^p\) and \(a_1=1,a_2=s\) \cite{rubensson2024}. The repeated-node values needed in the limit are \(f[1,1,1]=p(p-1)/2,\ f[1,1,0]=p-1,\ f[0,1,0]=1\). The determinant is \((p-1)h^2[h^2-(p-2)/2]<0\). Choose \(s>0\) and then \(\varepsilon>0\) small enough that both perturbed matrices are positive and their midpoint power difference \(D\) has \(\langle v|D|v\rangle<0\) for some unit vector \(v\).

Now let \(B=b P_v+(I-P_v)\), \(b>0\). As \(b\downarrow0\), \(\Tr(DB^{1-p})\to-\infty\), while the midpoint difference of the second term \(\Tr(B^p X^{1-p})\) stays bounded because the three choices of \(X\) are fixed and faithful. Thus the midpoint difference of \(F_p\) is negative for small enough \(b\). This disproves ordinary monotonicity. For the strong version, use processing Kraus operators \(I_4\otimes|0\rangle\langle+|\) and \(I_4\otimes|0\rangle\langle-|\) on the flag: both outcomes occur with probability \(1/2\) and yield \((A\oplus B)/T\) with an unoccupied flag sector. That sector is reducing for \(K\); all terms with nonzero coefficient are evaluated on the occupied faithful block. The same increase follows.
\end{proof}

As an explicit numerical check at \(p=3\), take \(A=\mathrm{diag}(16,1)\),
\(H=\left(\begin{smallmatrix}2&1\\1&0\end{smallmatrix}\right)\), \(\varepsilon=1\),
and \(B=I-\tfrac{31}{32}P_v\), \(v=(1,-4)^{\mathsf T}/\sqrt{17}\). Both
\(A\pm H\) are positive. Substitution in the midpoint formula gives an
input-minus-output difference of approximately \(-6.09708103\); the all-order
argument above does not rely on this numerical example.

\section{Equality at order two need not persist at interior orders}\label{app:equality}
An equality test at a single order requires care at \(p=2\). Here is a faithful example with fixed-Kraus flag erasure, in the same class of operations as Theorem~\ref{thm:five}. Take \(U=\mathrm{diag}(1,1,-1)\), and two positive matrices
\[
 A=\begin{pmatrix}1&1&0\\1&3&1\\0&1&2\end{pmatrix},\qquad
 B=\begin{pmatrix}2&2&0\\2&4&1\\0&1&2\end{pmatrix}.
\]
The reference is conjugation by \(U\) on the qutrit, acting trivially on the flag. The input is \((A\oplus B)/14\). The effective qutrit after flag erasure is \((A+B)/14\). For \(D=A,B,A+B\), one finds the same matrix \(DUD^{-1}\). Consequently \(f_2(D)=\Tr(D^2 U D^{-1}U)\) is additive on this pair: it is \(\Tr[(DUD^{-1})D(DUD^{-1})^\dagger]\). The values at \(A,B,A+B\) are \(34/3,40/3,74/3\), so \(I_2\) equals \(8/21\) both before and after.

For an interior order, the loss is a positive-weight integral of nonnegative differences for \(h_t(x)=(x-1)^2/(x+t)\), by \eqref{eq:beta} and the pinching/partial-trace argument (partial trace can be represented by a unitary twirl on the flag). Affine terms cancel, so vanishing at any one interior order would force equality of the resolvent differences at every \(t>0\) by continuity. Their large-\(t\) expansions would in turn give equality of all moments. But \(f_4(D)=\Tr(D^4 U D^{-3}U)\) takes values \(202,2210/9,11906/27\), respectively. The normalized fourth moment difference is \(89/189\), not zero. It is used only to detect unequal resolvents, not as a monotonicity statement at order four. Thus the loss is strict for every \(0<p<2\). This example instantiates, in the fixed-reference setting, the distinction between equality for interior orders and at the quadratic endpoint; see \cite{hiai2023equality,hiaimosonyi2017} for general quasi-entropy and divergence equality theory.

\end{document}